    \documentclass[final,3p,times,twocolumn]{elsarticle}

 \usepackage{graphicx}

\usepackage{amssymb}

\journal{Physics Letters B}

\begin{document}

\begin{frontmatter}



\title{$\alpha$-particle condensate states     in $^{16}$O}


\author[rvt]{S. Ohkubo\corref{cor1}} 
\ead{shigeo@cc.kochi-wu.ac.jp}
\author[focal1]{Y. Hirabayashi}   

\cortext[cor1]{Corresponding author}
\cortext[cor2]{Shigeo Ohkubo}

\address[rvt]{Department of Applied Science and Environment,
Kochi Women's University, Kochi 780-8515, Japan and\\
Research Center for Nuclear Physics, Osaka University, Ibaraki, 
Osaka 567-0047, Japan}

\address[focal1]{Information Initiative Center,
Hokkaido University, Sapporo 060-0811, Japan
}

\begin{abstract}
\par
 The  existence of a rotational band
 with the $\alpha$+$^{12}$C($0_2^+$)  cluster structure, in which   three $\alpha$ 
particles  in $^{12}$C($0_2^+$)  are locally  condensed,  is demonstrated  
 near  the four-$\alpha$ threshold of
 $^{16}$O in agreement with experiment.
This is achieved by  studying  structure    and  scattering for the 
  $\alpha$+$^{12}$C($0_2^+$)  system  in a unified way.  
 A drastic  reduction (quenching) of the moment of the inertia of the  $0^+$  state 
at 15.1 MeV just above the four-$\alpha$ threshold   in $^{16}$O suggests that it 
could be a candidate for the  superfluid state 
    in    $\alpha$-particle condensation.
\end{abstract}

\begin{keyword}
$\alpha$-particle condensation  \sep $\alpha$+$^{12}$C  scattering
\sep $\alpha$-cluster structure  \sep $^{16}$O \sep superfluid
\PACS  21.60.Gx \sep 25.55.Ci  \sep 27.20.+n  \sep 03.75.Nt


\end{keyword}
\end{frontmatter}



      $\alpha$-particle condensation has been 
 paid  much attention in  light  nuclei. 
     Up to now the     $0^+_2$ (7.65 MeV) Hoyle state
 of  $^{12}$C  has been considered to be a  candidate for a 
three $\alpha$-particle condensate  with a dilute density
 distribution \cite{Tohsaki2001}. The  elaborate microscopic 
 $\alpha$-cluster model wave functions of the Hoyle state  by  
 Uegaki  et al.
 \cite{Uegaki1979} and Kamimura  et al. \cite{Kamimura1981},
 which reproduce many experimental data involving the Hoyle state 
 are  almost completely equivalent  to the
    condensate wave function \cite{Funaki2003}.  
Many theoretical studies 
\cite{Tohsaki2001,Matsumura2004,Yamada2005,Funaki2005,Ohkubo2004,Ohkubo2007}  support the dilute
 property  of the Hoyle state.
However, the typical  physical modes 
such as  superfluidity and/or  a quantum vortex have not been observed.

 The fundamental question that may arise is that superfluidity due to 
    $\alpha$-particle condensation is
  difficult to  observe 
in    nuclear systems like $^{12}$C and  $^{16}$O, while  superfluidity has   
clearly been observed in  bulk systems such as He II and  $^3$He liquids.
We note, however,    that recent studies of parahydrogen 
\cite{parahydrogen}  and He clusters \cite{Hecluster}  show that superfluidity
can be observed  in small systems with 10  or less  particles.
This  encourages us to study the superfluidity of a small number of $\alpha$ particles 
 in  strong-interaction  systems composed of   protons and neutrons.  
 One of the most convincing   ways  to demonstrate  the existence 
of $\alpha$-particle condensation 
is to  show   superfluidity of the system. 
It has been shown    \cite{Stringari1996} theoretically
and experimentally   that a
 reduction (quenching)   of the moment of inertia from the rigid-body value is characteristic  to the 
superfluid behavior of a dilute Bose gas   due to the occurrence 
 of Bose-Einstein condensation. This reduction is also observed   for liquid helium 
(the Hess-Fairbank effect) \cite{Hess1967}. Path Integral Monte Carlo (PIMC) is a simulation which 
 makes use of   this reduction of the moment of inertia from the classical moment of inertia
  in calculating the 
superfluid density, for example, of   quantum fluids in confined geometries \cite{Ceperley2001}.

In  studies of $\alpha$-particle condensation in $^{16}$O, so far mostly  
   the $0^+$ state  has been discussed  \cite{Tohsaki2001,Wakasa2007,Funaki2008C}. 
Tohsaki   et al. \cite{Tohsaki2001} thought  
 that the  0$^+$ state at $E_x$=14.0 MeV located 
   below the four-$\alpha$ threshold energy   is an $\alpha$-particle  condensate. 
Wakasa  et al.   and  Funaki  et al.  \cite{Wakasa2007} suggested that    
a newly  observed 0$^+$ state at $E_x$=13.6 MeV  is an  $\alpha$-particle condensate. 
   On the other hand, very recently   Funaki  et al.
\cite{Funaki2008C}   performed a semi-microscopic four-$\alpha$ cluster model 
 calculation  in the OCM (Orthogonality Condition Model) and concluded that  the 0$^+$ 
state at  $E_x$=15.1  MeV  can probably be an $\alpha$-particle condensate. 
These  states were shown to have    
 dilute density distributions in the frame of the {\it bound state approximation}.
A dilute density distribution  is not equivalent to $\alpha$-particle condensation  and  
no clear  experimental  evidence for    $\alpha$-particle 
 condensation   such as  superfluidity and/or vortex excitation 
has  been  presented.  
$\alpha$-particle condensation  in $^{16}$O  has been   controversial. 
 We note that in experiment in the  high
 excitation energy region above the  four-$\alpha$
 threshold  well-developed $\alpha$-cluster  states ($2^+$, $4^+$ and
  $6^+$) have been  observed 
in the $^8$Be+$^8$Be and  $\alpha$+$^{12}$C($0^+_2$)  decay   channels of $^{16}$O 
  \cite{Chevallier1967,Freer1995,Freer2004,Freer2005}.
 These states have been  considered 
as   linear chain states of four $\alpha$ particles for many years 
\cite{Chevallier1967}.
We think that  it is important to understand not only the resonant  $0^+$ state
  but also the other resonant 
 higher spin states
 built  on it in a unified way in the context of $\alpha$-particle condensation.
It is also important to study the $\alpha$ decaying resonant  states above the
$\alpha$ threshold by solving  the {\it scattering equation} correctly.

In  this   paper, from a  unified description  of   structure 
  and scattering for the $\alpha$+$^{12}$C  system, we   show that a rotational 
band with the $\alpha$+$^{12}$C($0^+_2$) cluster structure  is   predicted near 
 the  four-$\alpha$  threshold in $^{16}$O. 
 The above    $\alpha$-cluster states observed in the four-$\alpha$ decay
 channel   \cite{Chevallier1967,Freer1995,Freer2004,Freer2005}
 can  be  understood consistently as  fragmented states of the band.
It is shown   that the  observed  $0^+$ state 
 at $E_x$=15.1 MeV  just above  the four-$\alpha$ threshold, which we  interpret
 to be fragmented  from the broad band head $0^+$ state,    has a reduced
 moment of inertia compared to   the   well-developed 
$\alpha$+$^{12}$C($0^+_2$)  cluster structure. It is suggested that this $0^+$ state 
could be a candidate for the  superfluid state in $^{16}$O 
in   four $\alpha$-particle condensation.

In  $\alpha$-cluster studies  a unified description of structure 
  and  scattering has been
   powerful because the  interaction potential can be uniquely determined
 from  rainbow scattering \cite{Khoa2007}.
 In fact, a unified study of  structure  
 and scattering of the $\alpha$+$^{40}$Ca system could disentangle a
 long-standing controversy about the $\alpha$-cluster structure in 
$^{44}$Ti \cite{Michel1986,Yamaya1998}.
This unification  
 may be extended to the case where a target nucleus is excited because 
 the nuclear rainbow  and prerainbow appear also 
 in inelastic scattering and the mechanism can be 
  understood in a similar way  to elastic scattering  \cite{Michel2004,Khoa2007}.

We  study the elastic and inelastic   
   $\alpha$+$^{12}$C scattering, and  states with the $\alpha$+$^{12}$C 
cluster structure in a unified way
  using a double folding (DF)  model 
 in the coupled channel method by taking into account 
 the excited   states of $^{12}$C, which has been  shown to be successful
in describing $\alpha$ and  $^3$He scattering at the high and low energy
regions \cite{Ohkubo2004,Ohkubo2007}. 
 The diagonal and coupling potentials of the DF model   for the $\alpha$-$^{12}$C system are
calculated as follows:
\begin{equation}
V_{ij}({\bf R})   = 
\int \rho_{00}^{\rm (\alpha)} ({\bf r}_{1})
     \rho_{ij}^{\rm (C)} ({\bf r}_{2})
v_{\rm NN} (E,\rho,{\bf r}_{1} + {\bf R} - {\bf r}_{2}) 
{\rm d}{\bf r}_{1} {\rm d}{\bf r}_{2},
\end{equation}
\noindent
where $\rho_{00}^{\rm (\alpha)} ({\bf r})$ is the ground state density
of the $\alpha$ particle, while $v_{\rm NN}$ denotes the density dependent
M3Y effective interaction (DDM3Y) \cite{Kobos1984} usually used in the DF model.
$\rho_{ij}^{\rm (C)} ({\bf r})$ represents the diagonal ($i=j$) or
transition ($i\neq j$) nucleon density of $^{12}$C
which is calculated using  the resonating group method by Kamimura  et al. 
\cite{Kamimura1981}.
In the calculation of densities of $^{12}$C,
the shell-like structure of the ground state 0$^+_1$ 
  and the well-developed $\alpha$-cluster structure of the Hoyle 0$^+_2$ 
    state are simultaneously well   reproduced. 
These wave functions 
    have  been checked  against  many experimental data including charge
     form factors, electric transition probabilities and 
     reactions involving      excitation to the 0$^+_2$  state \cite{Kamimura1981}. 
The  wave function for the  
   Hoyle  0$^+_2$ state is very  close   to the $\alpha$-condensate wave function.
In the calculations the normalization factor  $N_R+iN_I$ is introduced
 for the  $\alpha$-$^{12}$C  DF  potential. The  imaginary part takes into
 account   absorption phenomenologically.
 
%
\begin{figure}[tbh]
   \includegraphics[width=6.8cm]{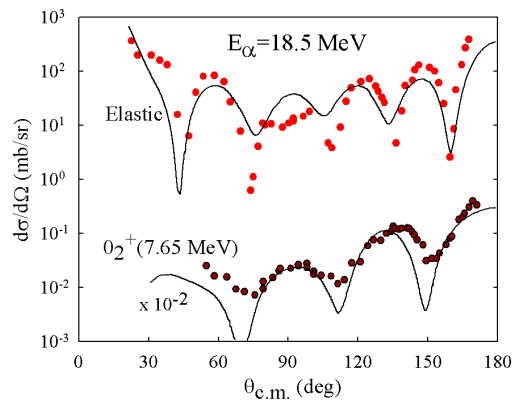}

 \protect\caption{\label{fig.1} { Calculated angular distributions 
(solid lines)  in  elastic and inelastic $\alpha$+$^{12}$C   
scattering  at $E_\alpha$=18.5 MeV are compared with the  experimental data (points) \cite{Mitchell1964}.   }
}
 \end{figure}

 We have shown in Ref.\cite{Ohkubo2004} that elastic and
  inelastic  $\alpha$+$^{12}$C rainbow scattering  from the Hoyle state   
 in the  high 
energy region ($E_\alpha$=139, 166 and 172.5 MeV) can be well described  
 in the  DF   model with $N_R$=1.23-1.26.
 We extend this analysis  to the  lowest energy  $E_\alpha$=18.5 MeV 
 where experimental data of both  
 inelastic scattering from the Hoyle state 
and elastic scattering are available \cite{Mitchell1964}. 
We analyze the angular distributions in the coupled channel calculations 
 including  most of   the  channels of $^{12}$C  open at this energy, that is,
  g.s, 2$_1^+$(4.44 MeV), 3$^-$(9.65 MeV),  0$^+_2$(7.65 MeV) and 2$^+_2$(10.3 MeV). 
The absorption due to the coupling to   all the other open channels, i.e., 
 $p$+$^{15}$N, $n$+$^{15}$O 
and  $d$+$^{14}$N channels, is introduced as a small imaginary potential with $N_I$=0.045.  
The calculated angular distributions  with $N_R$=1.398 are compared
with the experimental data  in Fig.~1. 
 The volume integral per nucleon pair for the real potential is $J_V$=427.3 MeVfm$^3$ for the
 g.s. channel.
 The characteristic    oscillations at the backward 
hemisphere of the experimental data are well reproduced  by the calculation.
The backward rise in elastic scattering, ALAS (Anomalous Large Angle Scattering), is     
 caused by the internal waves \cite{Brink1985}, which penetrate  deep into the internal 
region of the potential. 
 The ALAS  seen for inelastic scattering  from
the Hoyle state  is also understood  similarly in terms of internal 
 waves  \cite{Michel2004}. 
  These results suggest   that the  diagonal and  coupling potentials
in Eq.~(1)   work 
in the  $\alpha$-cluster  structure study in the  low energy region near the
$\alpha$+$^{12}$C($0_2^+$) threshold.

We study the  resonant $\alpha$-cluster structure  of 
 $^{16}$O  by solving the
 coupled channel scattering equations with use of the real part of the  double 
folding potential. 
  We take $N_R$=1.34, which is chosen 
so that the calculated energy of the band head  $1^-$ state of the 
$K=0_1^-$,  which has  a well-developed  $\alpha$+$^{12}$C(g.s.)
  structure, corresponds  well with  experimental energy.
The volume integral $J_V$ and   rms radius of the potential are  409.6   MeVfm$^3$ 
and 3.44 fm for the g.s. channel, and  510.3 MeVfm$^3$ and 4.29 fm for
 the $\alpha$-$^{12}$C($0_2^+$)  channel, respectively.
The energy-dependence of $N_R$, that is,  $N_R$=1.23 at
 $E_\alpha$=139 MeV \cite{Ohkubo2004}, $N_R$=1.389 
at $E_\alpha$=18.5 MeV
 and  $N_R$=1.34 at $E_\alpha$$\simeq$0, which increases  toward the lower energy  as
 the incident energy  decreases and 
 again increases toward   zero energy, seems to be  consistent
  with the threshold anomaly. 
 
\begin{figure}[tbh]
   \includegraphics[width=8cm]{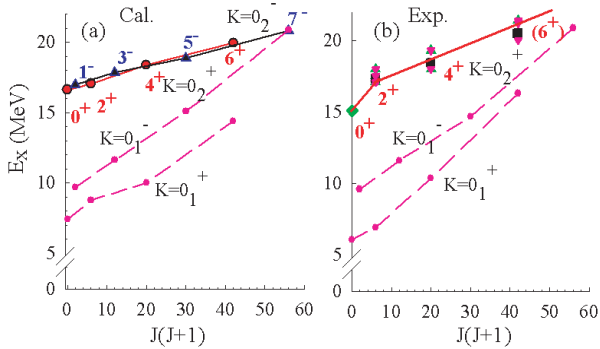}
   \protect\caption{\label{fig.2} { (a) Calculated  states  
of  the $K=0^+_1$ and   $K=0^-_1$ parity-doublet bands with the $\alpha$+$^{12}$C(g.s)  
structure, and the $K=0^+_2$ and $K=0^-_2$  bands with the 
$\alpha$+$^{12}$C($0^+_2$)  cluster  structure   near the four-$\alpha$ threshold 
  are compared  with  (b) the experimental  levels in   $^{16}$O taken from 
   Ref.\cite{Chevallier1967} (triangle up), Ref.\cite{Freer1995,Freer2004,Freer2005} (triangle down) and 
 Ref.\cite{Tilley1993} (diamond). The centroid of each of the spin states is  shown by a
 square.  The lines are   to guide the eye.
    }
}
\end{figure}

In Fig~2 the calculated  states are shown  in comparison with the 
experimental data. 
  The calculation reproduces the parity-doublet  $K=0_1^+$ band and 
 $K=0_1^-$  band with the 
 $\alpha$+$^{12}$C(g.s.)  structure well.
A resonance energy $E_r$ is defined as the energy where the elastic channel 
phase shift $\delta_J$ passes $\pi$/2.
(The $0^+$  state of the $K=0_1^+$  band is  well below the barrier and
 calculated in the single channel bound state approximation.) 
    In Table I the    calculated excitation energy of resonant states of 
the  parity-doublet  
$K=0^+_1$ and   $K=0^-_1$  bands 
 and its widths derived from 
$\Gamma_\alpha=2/\frac{d \delta_J}{d E}\Bigm|_{E=E_{r}}$
are displayed
in comparison with  the  experimental  data \cite{Tilley1993}.
Because the calculated widths  are strongly  dependent on excitation energy,
  the 
dimensionless reduced widths $\theta^2_\alpha$, which are more physically related to the 
  degree of $\alpha$-clustering,  are also shown at the three channel radii
$a$=5.2, 5.6 and 6.0 fm.
  $\theta^2_\alpha$ is defined as   $\Gamma_\alpha$=2$P(a)\gamma^2 (a)$
 with $P(a)$ being the penetration factor,
 $\gamma^2 (a)=\theta^2_{\alpha}(a)\gamma^2_w (a)$ and the
  Wigner limit  $\gamma^2_w (a)$=3$\hbar^2/2\mu a$  with $\mu$   denoting  
the reduced mass.   The agreement of the calculated 
 $\theta^2_\alpha$ with
experiment is good especially for the  $K=0_1^-$ band states. 
The large $\theta^2_\alpha$  reconfirms that the 
  $K=0_1^-$ band    has a well-developed $\alpha$-cluster structure.
The rather small calculated reduced widths for the positive parity  states may be 
 improved  by introducing a small  parity-dependence in  the potential 
in order to reproduce the experimental excitation energy
 \cite{Baldock1984}.

\begin{table*}[th]

\begin{center}
 \setlength{\tabcolsep}{7pt}
  \footnotesize
 
\caption{ The calculated excitation  energy,  $\alpha$-decay width   $\Gamma_\alpha$ 
 and  dimensionless reduced
 width  $\theta^2_\alpha(a) $ of the  resonant states of the 
$K=0^+_1$ and   $K=0^-_1$  bands     
are compared with the experimental data \cite{Tilley1993}.
$\theta^2_\alpha(a) $ is evaluated at the channel radii   $a$=5.2, 5.6 and 6.0 fm.
}
\begin{tabular}{lccccccccccc}
 \hline \hline  
    &              & \multicolumn{5}{c}{cal.}                &       \multicolumn{5}{c}{exp.}      \\                             
                &  $J^\pi$ &   $E_x$ &  $ \Gamma_\alpha$  &    &{$\theta^2_\alpha(a)$}  &  
              &                   $E_x$        &  $\Gamma_\alpha$    &   &   {$\theta^2_\alpha(a)$}  &    \\
              &                    &  (MeV)   &      (keV)                       &   &     $a $ (fm)                 &       
              &                   (MeV)         &      (keV)                       &   &   $a$  (fm)                             &  \\
                    &       &             &           & 5.2 &  5.6 &  6.0 &                              &         &5.2 &  5.6 &    6.0 \\
\hline
$K=0^+_1$  & $4^+$    &    10.00     & 5       & 0.15 & 0.09  &  0.06  &   10.36         &  26$\pm$3  &  0.35  & 0.22  &  0.14       \\       
                     & $6^+$    &  14.37         &  34   & 0.10 &  0.06 & 0.04  &  16.28         &  420$\pm$20  & 0.36 &  0.26 & 0.20                                \\
   $K=0^-_1$  &   $1^-$    &  9.67           & 778   & 1.02  &  0.87 &  0.79   &  9.59            & 480$\pm$20  &  0.71 & 0.61 & 0.54                         \\    
                    &   $3^-$    &  11.61         &  822  &0.64 & 0.54 & 0.48        &  11.60          &800$\pm$100    & 0.63 & 0.53 & 0.47                \\    
                    &   $5^-$    &  15.09        &    440   & 0.26 & 0.21 & 0.18    &  14.66           &  672$\pm$11    & 0.49 &  0.38 & 0.31               \\     
                   &   $7^-$        & 20.86        &  1790   & 0.77 &  0.59 & 0.49 &  20.86           &  904$\pm$55   & 0.39 & 0.30 & 0.25              \\
      \hline                      
 \hline                          
\end{tabular}
\end{center}
\end{table*}

In  Fig.~2  the $\alpha$-cluster structure, in which $^{12}$C is excited to the
 Hoyle state, is shown.  These resonant states with  positive parity form 
 a $K=0_2^+$ band structure. 
 The resonant structure with the $\alpha$+$^{12}$C($0^+_2$) is   seen 
 by investigating the S-matrix  and partial wave cross sections from the ground state to 
the Hoyle state.  
 The modulus of the 
S-matrix S$_{g.s.-0_2^+}$, which indicates the transition strength of the incident flux 
to the
 Hoyle state, shows a peak at the energy which corresponds to a resonant  state
 of the composite $\alpha$+$^{12}$C($0_2^+$)  system.  Accordingly the partial wave 
 cross section also shows a peak at the energy  corresponding to the peak of the S-matrix. 
In Fig.~3(a) the partial cross sections  calculated 
in  the  coupled five   channels    are 
displayed. In Fig.~3(b) the partial cross sections 
   calculated in the
  two channels (g.s. and  $0_2^+$)  are  shown to help to understand the origin of 
the peaks. 
In the two channel calculations   prominent peaks are seen  clearly,
which corresponds to a resonance of the composite system  with the 
$\alpha$+$^{12}$C($0_2^+$) cluster structure.
In the five channel calculations each  peak is  fragmented.
The resonant structure with the $\alpha$+$^{12}$C($0_2^+$) cluster $K=0_2^-$ band 
 is also obtained for negative parity states.   
In Table II the excitation energy of the  calculated  resonant states 
of the parity-doublet  $K=0_2^+$ and 
 $K=0_2^-$ bands 
and  its  widths are displayed. These  are  evaluated 
 from the partial cross sections from the ground state to the $0_2^+$ state assuming a 
 Breit-Wigner function. 
As seen in Fig.~2, quite remarkably  our potential, which reproduces the higher energy rainbow scattering 
and low energy
scattering and the lowest $\alpha$-cluster states of the parity-doublet  $K=0^+_1$ and 
$K=0^-_1$ bands,   locates the $K=0^+_2$  band with the $\alpha$+$^{12}$C($0^+_2$) cluster
 structure near  the $\alpha$+$^{12}$C($0^+_2$)  threshold ($E_x$=14.82 MeV), namely 
 near the four-$\alpha$  threshold ($E_x$=14.44 MeV).
 It is noted  that our predicted  $K=0^+_2$ band states as well as the
 $K=0^-_2$ band states 
are on the rotational $J(J+1)$ trajectory. 
   The Pauli principle is not important in this highly excited
 energy region. As for the $K=0^-_2$ band, the observed states, $1^-$ at $E_x$=16.20 MeV
with $\Gamma$=0.58 MeV and    $5^-$ at $E_x$=18.40 MeV with $\Gamma$=0.55 MeV could be a 
candidate.  No   $3^-$  state has been reported at the corresponding excitation energy.
More experimental studies are needed for the negative parity states 
including the coincidence experiments.
  We have investigated the  $N_R$  dependence of the  location
  of the $K=0^+_2$ band.  In  calculations
 the  band head energies are  
  $E_x$=16.9,  16.6,  and 16.9 MeV for $N_R$=1.30,  1.34,
   and 1.398, respectively, 
  which shows that the  $N_R$-dependence is mild  in the relevant region.

\begin{figure}[tbh]
   \includegraphics[width=5.8cm]{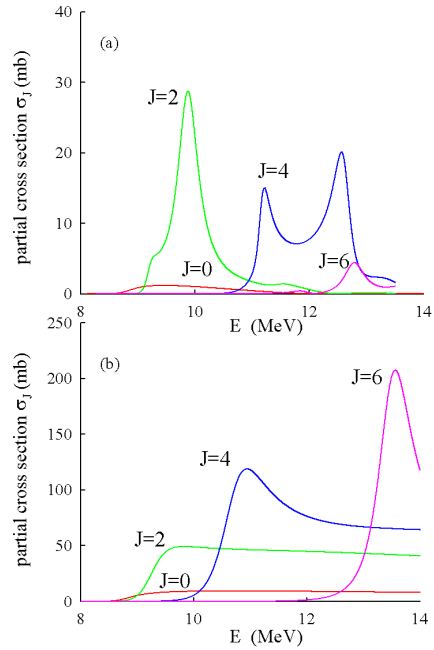}
   \protect\caption{\label{fig.3} {
Partial wave cross sections    in inelastic scattering from the Hoyle state 
 in the coupled channel calculations are displayed as a function of the incident 
energy of the $\alpha$ particle in the center of mass system $E$,  
 (a) coupled  five channels (g.s., $2_1^+$, $3^-$, $0_2^+$ and $2_2^+$) and 
(b) coupled  two channels (g.s. and  $0_2^+$).
   The vertical scales of (a) and (b) are different.
    }
}
\end{figure}

\begin{table}[th]
\begin{center}
\caption{ The calculated excitation energies and  widths of the   $K=0^+_2$ and
  $K=0^-_2$ 
 band states.
}
\begin{tabular}{cccccc}
 \hline
  \hline
\multicolumn{3}{c}{$K=0^+_2 $ } & \multicolumn{3}{c}{$K=0^-_2 $}     \\           
        $J^\pi$ &   $E_x$ &  $ \Gamma$              &$J^\pi$ &   $E_x$        &  $\Gamma$      \\
                           & (MeV)       & (MeV)              &              & (MeV)       & (MeV)   \\   
 \hline
   $0^+$  & 16.61 &    1.14 &     $1^-$  & 16.98 &   0.57            \\
   $2^+$    &  17.04 &  0.45 &                      $3^-$  &  17.83 &   0.27            \\
   $4^+$    & 18.38 &  0.23 &                    $5^-$  &  18.86 &   0.24            \\
   $6^+$    &  19.95 &  0.39 &                    $7^-$ &  20.81 & 1.17                \\

 \hline                          
 \hline                          
\end{tabular}
\end{center}
\end{table}

%
 The  $2^+$, $4^+$ and   $6^+$    states  of the $K=0^+_2$ band shown in  Fig.~2(b) were observed in the 
 coincidence experiments to search for the four $\alpha$-particle states   
 \cite{Chevallier1967,Freer1995,Freer2004,Freer2005}. 
 In more detail, Chevallier  et al. \cite{Chevallier1967} first  observed the   $2^+$ (16.95 MeV), 
 $2^+$ (17.15 MeV), $4^+$ (18.05 MeV) and  $6^+$ (19.35 MeV) states  in the
 $^{12}$C($\alpha$,$^8$Be) reaction, which is  interpreted to  lie on the rotational
 band with a very small  rotational constant $k$=${\hbar^2}$/${2 {\cal I}}$=64 keV 
where ${\cal I}$ is   the moment of inertia.  
They concluded from this very large moment of inertia  
 that  ``the only conceivable structure 
with such a moment of inertia is of four $\alpha$'s laid out in a string and 
rotating rigidly" and thereafter the band has been considered to be a linear chain 
for more than four decades.  
Later, Freer   et al.   performed the $^{12}$C($^{16}$O,4$\alpha$) reaction \cite{Freer1995} and observed 
  the  two $2^+$ (17.1  and 17.5 MeV), $2^+$ or  $4^+$ (18.0 MeV), $4^+$ (19.3 MeV) and 
 ($6^+$) (21.4 MeV) states, which they interpreted to lie  on the rotational
 band with a  small   $k$=95$\pm$20 keV.  Furthermore  Freer   et al.  observed the high spin states,  
     two $6^+$ states at 20.0 and 21.2 MeV in the
 $^{12}$C($^{12}$C,$^{8}$Be+$^{8}$Be) reaction \cite{Freer2004,Freer2005}.
  We interpret that the $\alpha$ strength of all the  above   states observed
 in the coincidence experiments 
  may be considered to be fragmented from a well-developed 
cluster band and the 
 centroid of each of the spin states is  displayed in Fig.~2(b).   
 (The fragmentation of the $\alpha$ strength is caused by the coupling between the 
 genuine cluster band state and other states nearby, which  
 has been demonstrated experimentally and theoretically  in the typical
 $\alpha$-cluster structure 
  in   $^{40}$Ca and      $^{44}$Ti  \cite{Yamaya1998,Sakuda1998}.)
The  rotational constant estimated   from each  centroid of the $2^+$, $4^+$ and  $6^+$ states 
is $k$=86 keV, which    
is also very small consistent with the above values  by  Freer   et al.  \cite{Freer1995} and
   Chevallier  et al.  \cite{Chevallier1967}. 
The very large moment of inertia, about four times that  of the $K=0^+_1$  
band built  on the $0^+$ state at 6.06 MeV, which has an $\alpha$+$^{12}$C(g.s.)  cluster 
 structure, implies a very extended structure of $^{16}$O in these states. 
Our  calculated        $k$=80 keV 
  is  close to the experimental value $k$=86 keV.
The large  moment of inertia is caused by the mechanism that   the 
 $\alpha$ 
particles are locally condensed in the  Hoyle state with an  extended density 
distribution. 
The moment of inertia \cite{Goldstein1964} classically 
calculated from the matter distribution used in Eq.~(1)  shows that 
   56\% of     the large total moment of inertia,   
  which is given by  
${\cal  I}={\cal  I}({\alpha})+{\cal  I}(^{12}C(0_2^+))+{\cal  I}$(rel.) where 
 ${\cal  I}({\alpha}$),  ${\cal  I}(^{12}$C$(0_2^+))$ and  ${\cal  I}$(rel.) stand for
 the moment of inertia of the  $\alpha$ particle, $^{12}$C$(0_2^+)$ and
  their relative motion, respectively, comes from ${\cal  I}(^{12}$C$(0_2^+))$
, i.e.,  the   extended $\alpha$-particle  distribution in the Hoyle state.
   40\% of ${\cal  I}$  comes   
   from ${\cal I}$(rel.), from which  the intercluster distance $R$ between the orbiting $\alpha$ particle 
and the $^{12}$C($0^+_2$) core  is estimated to be  $R$=5.9 fm. 
Considering that the superfluid part of  $^{12}$C$(0_2^+)
$ due to 
$\alpha$-particle condensation, whose probability is about 70\% 
\cite{Matsumura2004,Yamada2005}, does not
contribute to   ${\cal I}(^{12}$C$(0_2^+))$,  the 
${\cal  I}$(rel.) becomes larger, that is,   the intercluster distance increases significantly. 
 If we write the total moment of inertia as ${\cal  I}$=$\mu$$R^2$, we  get
 $R$=9.2 fm,
 which  is very large,  although  smaller than 12.3 fm in the $\alpha$-particle  linear chain  
picture given in Ref.\cite{Chevallier1967},  suggesting  an 
  $\alpha$-particle halo configuration. 
It is very interesting and important to measure the 
size of the states  of the $K=0^+_2$  band experimentally.
The fact  that  the member states, the two $2^+$ (17.1  and 17.5 MeV),  
($2^+$ or  $4^+$ ) (18.0 MeV) 
 and  $4^+$ (19.3 MeV) states, are  observed in the $\alpha$+$^{12}$C($0^+_2$)  decay channel
 \cite{Freer1995} are also reasonably understood by  our picture  that the band has an 
 $\alpha$+$^{12}$C($0^+_2$)   cluster structure instead of the linear chain structure
of the four $\alpha$ particles.
 Since the channel coupling
 between the  Hoyle state and the 
2$^+_2$ state,  which has a well-developed $\alpha$+$^8$Be  cluster structure as well 
as the  Hoyle state \cite{Uegaki1979}, is strong in the present calculation, the   $^8$Be+$^8$Be  configuration
 might also  contribute to the deformation. The fact that  the band states are 
observed in 
the  $^{12}$C($\alpha$,$^8$Be) reaction seems to be  consistent with this possibility.

What is especially interesting  is the $0^+$ state.
 Our calculation locates the band head $0^+$ state at around 16.6 MeV.
 However, in the above-mentioned coincidence experiments no $0^+$  state  was 
 observed at the predicted  energy region. 
 In the compilation of the energy levels of  $^{16}$O in Refs.\cite{Tilley1993,Ajzenberg1977}
 no $0^+$ state  with 
a large  $\alpha$-decay width has been observed  near  16.6 MeV.
(A state reported at 16.36 MeV ($0^+$, $1^-$) in Ref.\cite{Ames1982}, which could be 
 the same state as the $0^+$ state at 16.33 MeV observed in the  (p,t) reaction \cite{Ajzenberg1977},
 has a very small $\alpha$-decay width of $\Gamma_\alpha$/$\Gamma$=0.07.)
The  reported states with a considerable $\alpha$ width 
in Refs.\cite{Tilley1993,Ajzenberg1977}
observed in $\alpha$-particle scattering from  $^{12}$C
  are  15.1 MeV ($0^+$), 17.6 MeV ($0^+$ or $1^-$), 
 and 18.1 MeV ( $0^+$, spin  tentative) with 
 $\Gamma_\alpha$/$\Gamma$=0.35, 0.32, and 0.31, respectively. 
 Because the  calculated band head $0^+$ state is very broad  with
  no centrifugal barrier as seen in Fig.~3(b), it  
may have escaped 
from  detection. The above observed states nearby may be considered to be fragmented
from the band head  $0^+$ state. The centroid of the above three  states is 16.9 MeV,
 which is close to our  predicted energy.

If    a four-$\alpha$ condensate $0^+$ state exists near the threshold,
it will be located below the calculated band head $0^+$ state with the 
$\alpha$+$^{12}$C($0^+_2$) configuration because the moment of inertia is reduced.
  As seen in Fig.~2(b), the moment of inertia of the  $0^+$ state  at 15.1 MeV  with 
a large $\alpha$ width     below 
 the  calculated band head energy and just above the four-$\alpha$ threshold 
is  drastically reduced to a quarter   of that of the $2^+$ and $4^+$  states 
of the $K=0^+_2$ band.  
Because the orbiting $\alpha$ particle is  also sitting in the $0s$ state like   
   the  three
$\alpha$ particles in the $0s$ state in $^{12}$C($0^+_2$), this  could
 be a good  candidate
 state in which four $\alpha$ particles are condensed   
spherically in the lowest $0s$ state.
  The probability of condensation is  roughly  
conjectured  to be three quarters  from the reduction of the moment of inertia.
 This is consistent with a recent calculation by Funaki et al. \cite{Funaki2008C}
 who claim that the probability
 that the four $\alpha$ particles are sitting in the $0s$ state is  61$\%$.
This $0^+$ state  is less dilute than the deformed $K=0_2^+$ band states.
The excited  $2^+$, $4^+$ and $6^+$ states of the $K=0_2^+$ band are created by lifting
 an  $\alpha$ particle  to the  $D$,  $G$ and  $I$ states, respectively,
 which brings about  the deformed $\alpha$+$^{12}$C($0^+_2$) configuration.
Similar excited states are expected in $^{12}$C. In fact, the observed $2_2^+$ state
has been discussed to be an   $\alpha$ condensate state \cite{Yamada2005,Funaki2005}.
However, a $4_2^+$ state and a  rotational band built on the Hoyle state have not been
 observed in $^{12}$C. 

Next we discuss  possible $\alpha$-particle condensation in the $^{16}$O$\sim$$^{20}$Ne region.
A  weak coupling  holds in the  $\alpha$-cluster structure 
 in the $^{16}$O$\sim$$^{20}$Ne region \cite{Hiura1972}.   The weak coupling works for the
states with a more developed cluster structure like the  $K=0_2^+$ band in $^{16}$O. 
Therefore a  band  with the X+$^{12}$C(0$_2^+$) cluster structure such as  $^8$Be+$^{12}$C(0$_2^+$), 
   analogous  to the  $K=0_2^+$ band in $^{16}$O, may be  expected near
 the $^{12}$C(0$_2^+$) threshold in $^{20}$Ne, $^{19}$F, $^{18}$O, and $^{17}$O. 
It is interesting to study whether  the  two fragmented  $10^+$ states  at $E_x$=35.2 and 36.5 MeV in $^{20}$Ne observed
in the $^8$Be+$^{12}$C(0$_2^+$) decay channel \cite{Freer2005} could be such
 an analogous rotational band state.  
The local condensation of $\alpha$ particles like  the  $^{12}$C(g.s.)+$^{12}$C(0$_2^+$) and 
$^{12}$C(0$_2^+$)+$^{12}$C(0$_2^+$)
  cluster structures as well as six $\alpha$-particle condensation in $^{24}$Mg is also  
  interesting.

The idea of   (local)  $\alpha$-particle condensation  
may be extended to heavier nuclei. 
 Yamada  et al. \cite{Yamada2004} discussed   possible  $\alpha$-particle
 condensation in  nuclei not heavier than $^{40}$Ca by solving the  Gross-Pitaevski  
and Hill-Wheeler equations. 
Gridnev   et al.  \cite{Gridnev2000} discussed  the fragmented 
$\alpha$-cluster states in $^{32}$S
observed in  ALAS  of $\alpha$ particles from 
$^{28}$Si    from the viewpoint of  $\alpha$-particle condensation. 
Ogloblin \cite{Ogloblin2006} and Oertzen \cite{Oertzen2006,Kokalova2006} 
 suggested  an $\alpha$-particle condensate in heavier  nuclei.    Local $\alpha$-particle condensation 
 will  not  be limited to $^{16}$O and might be  anticipated in other heavier nuclei including the 
 A=212 region. The detection  method  proposed in Ref.\cite{Kokalova2006} may be  useful.

To   summarize, in agreement with experiment we have shown that a rotational band with 
a developed
 $\alpha$+$^{12}$C($0^+_2$) cluster structure, in which three
 $\alpha$ particles are locally condensed, is  located  near the  
four-$\alpha$ threshold. This was achieved by using  the  double folding  potential which reproduces elastic 
and inelastic $\alpha$-particle scattering  from $^{12}$C.
 The moment of inertia of the  $0^+$ state at 15.1 MeV just above  the four-$\alpha$ threshold 
is drastically reduced  suggesting 
that this state could be a candidate for the superfluid state  of $\alpha$-particle 
condensation in $^{16}$O.
 It would also be  interesting to study 
 the local condensation of di-neutrons, deuterons and even $^{16}$O  in heavier nuclei.
 
 S.~O. would like to thank Profs.  S.~Koh and K.~Iida for
 useful communications.  This work has been supported by the Yukawa Institute for 
 Theoretical Physics.

\end{document}